\documentclass[%
amsmath,amssymb,
aps,
pra,
]{revtex4-2}

\usepackage{graphicx}
\usepackage{amssymb,amsmath}
\usepackage[utf8]{inputenc}
\usepackage{multirow}
\usepackage{xcolor}
\usepackage{hyperref}
\hypersetup{colorlinks=true,linkcolor=blue,anchorcolor=blue,citecolor=magenta,filecolor=blue,urlcolor=blue,bookmarksnumbered=true}
\usepackage[flushleft]{threeparttable}
\usepackage{mathtools}

\usepackage{amsmath}

\usepackage{dsfont}
\usepackage{amsthm}

\usepackage{enumitem}
\setlist[itemize]{leftmargin=10mm}

\usepackage{braket}

\begin{document}


\title{Systematic many-fermion Hamiltonian input scheme \\
	and spectral calculations on quantum computers} 

\author{Weijie Du\textsuperscript{1,2}}
\email[Email: ]{duweigy@gmail.com}

\author{James P. Vary\textsuperscript{2}}

\affiliation{\textsuperscript{1}Department of Physics and Astronomy, Iowa State University, Ames, Iowa 50010, USA}
\affiliation{\textsuperscript{2}Institute of Modern Physics, Chinese Academy of Sciences, Lanzhou 730000, China}

\date{\today}

\begin{abstract}
	
	We present a novel input scheme for general second-quantized Hamiltonians of relativistic or non-relativistic many-fermion systems.
	This input scheme incorporates the fermionic anticommutation relations, particle number variations, and respects the symmetries of the Hamiltonian. 
	Based on our input scheme, we propose a hybrid quantum-classical framework for spectral calculations on future quantum hardwares.
	We provide explicit circuit designs and the associated gate cost. 
	We demonstrate our hybrid framework by solving the low-lying spectra of ${^{42}}$Ca and ${^{46}}$Ca.
	Our input scheme provides new pathways to solving the spectra and dynamics of the relativistic and nonrelativistic many-fermion systems via first-principles approaches.

\end{abstract}

\maketitle



\

{\bf Introduction.}
Quantum computing has the potential to solve the structure and dynamics of many-body problems that are intractable with classical computers \cite{feynman1982simulating,lloyd1996universal} in the fields such as quantum chemistry \cite{cao2019quantum,McArdle_2020,bauer2020quantum,Beck:2023xhh,robledomoreno2024chemistryexactsolutionsquantumcentric,low2025fastquantumsimulationelectronic}, nuclear physics \cite{PhysRevLett.120.210501,PhysRevC.106.034325,PhysRevC.105.064317,Du:2020glq,Turro:2023dhg,Du:2023bpw,Wang:2024scd,Singh:2025ubp,Zhang:2024rpa,Turro:2024ksf}, elementary particles physics \cite{Kreshchuk:2020aiq,Kreshchuk:2023btr,Bauer:2023ujy, Bauer:2023qgm, Bauer:2025nzf, PRXQuantum.5.037001}, field theories \cite{Jordan_2012,PhysRevA.98.032331, jordan2019quantum, PhysRevD.101.074512,Kreshchuk:2020kcz,Kreshchuk:2020dla,PhysRevD.106.054508,bauer2022quantum,PhysRevLett.131.081901,PRXQuantum.4.030323,Davoudi:2024wyv,Abbott:2023coj,Qian:2024gph,Rhodes:2024zbr,Kane:2023jdo,Du:2024ixj,Wu:2024adk,Ciavarella:2025zqf,Burbano:2024uvn,Chen:2024pee,Barata:2025hgx,Qian:2025fnx}, and condensed matter physics \cite{PhysRevA.92.062318,PRXQuantum.2.030307,PhysRevResearch.4.023097,Kokcu:2023vwg, alipanah2025quantumdynamicssimulationadvectiondiffusion, mootz2024adaptivevariationalquantumcomputing, Getelina_2024, chen2024minimallyentangledtypicalthermal, alvertis2024classicalbenchmarksvariationalquantum}. One of the key questions in solving the structure and dynamics of relativistic and nonrelativistic many-body theories is: how to design a systematic Hamiltonian input scheme that is able to treat particle number variations and respects the symmetries of the underlying theory?

We present a novel Hamiltonian input scheme applicable to both relativistic and non-relativistic many-fermion systems in this work. 
This input scheme does not rely on the bookkeeping mapping between the fermion operators and the qubit operations such as the Jordan–Wigner transformation \cite{JordanWigner1928, PhysRevLett.79.2586} and the Bravyi–Kitaev transformation \cite{Bravyi_2002,BK_encodingScheme}, where each string of fermion operators in the second-quantized Hamiltonian is decomposed into a set of Pauli strings --- this decomposition becomes complicated when many-body interactions are involved. 
Instead, we create the quantum walk states based on the Hamiltonian and Fock states of the many-fermion system in second quantization.
We then project the walk states to the Hilbert subspace that encodes the Hamiltonian by pruning the walk-state components according to the symmetries of the system.
Unlike the sparse-matrix input scheme \cite{PhysRevLett.102.180501,childs2010relationship,berry2012black,gilyen2019quantum} that accesses the matrix element based on row and column indices, our input scheme restores the natural connections between the Hamiltonian matrix and Fock states.
It also dynamically computes the many-body matrix elements from the few-body matrix elements in a way that incorporates the Pauli principle, anticommutation relations, and particle number variations that may be present.
In this work, we provide a systematic and practical circuit design of our input scheme. 

Based on our input scheme, we also propose the framework of the symmetry-adapted (SA) quantum Krylov subspace diagonalization (QKSD) method for the spectral solutions of many-fermion systems.
In particular, our input scheme enables straightforward generations of the Krylov bases \cite{watkins2007matrix, Kirby_2023,Kirby_2024, motta2024subspace} with selected symmetries.
In the resulting SA Krylov subspaces, we can evaluate the many-fermion matrix elements on the quantum computer. 
These elements are then input to the classical computer for solving the eigenenergies of states with elected symmetries via a greatly reduced matrix eigenvalue problem. 
While the standard QKSD methods \cite{Huggins_2020,doi:10.1021/acs.jctc.9b01125,PhysRevA.105.022417,PRXQuantum.3.020323,doi:10.1137/21M145954X,PhysRevA.107.032414,Kirby_2023,Kirby_2024, motta2024subspace} are only capable of solving the ground-state eigenenergy, the symmetry control over the Krylov subspaces facilitates excluding low-lying states with undesired symmetries and extending our reach to excited states. 

We demonstrate our many-fermion Hamiltonian input scheme and the hybrid quantum-classical SA-QKSD framework via simple nuclear structure calculations, where we compute the low-lying spectra of the Calcium isotopes. Compared to the conventional variational quantum eigensolver \cite{peruzzo2014variational,mcclean2016theory} where the measurement overhead and optimization are in general challenging, our hybrid framework does not require extensive parameter optimizations and enables access to excited states as well. 
In addition to the spectroscopy, we can also implement our input scheme with efficient Hamiltonian simulation algorithms \cite{PhysRevLett.118.010501,gilyen2019quantum,Berry2020timedependent} for dynamics study.

{\bf Many-fermion Hamiltonian.}
The general Hamiltonian $H$ of a many-fermion system with possible particle number variations is
\begin{equation}
	H = \sum _j \langle Q_j |H | P_j \rangle b^{\dagger}_{Q_j} b_{P_j},
	\label{eq:labeling_i}
\end{equation}
with the mappings $Q_j \mapsto \{ p_j, q_j, \cdots , r_j  \} $ and $P_j \mapsto \{ u_j, v_j, \cdots , w_j \}$ such that the few-body matrix element $ \langle Q_j |H | P_j \rangle = \langle p_j q_j \cdots r_j | H | u_j v_j \cdots w_j \rangle$, $ b^{\dagger}_{Q_j} = a_{p_j}^{\dagger} a_{q_j}^{\dagger} \cdots a_{r_j}^{\dagger} $, $ b_{P_j} = a_{w_j} \cdots a_{v_j} a_{u_j}  $, and $j\in [0, \mathcal{D}-1]$ runs over all possible ``P-Q" pairs for which matrix elements of $H$ are non-vanishing. $ H$ can contain terms which change the particle number. The anti-commutation relations hold for the fermion operators, i.e.,  $ \{ a^{\dagger}_p, a_q \}=\delta _{pq} $, and $ \{ a^{\dagger}_p, a^{\dagger}_q \} = \{ a_p, a_q \} = 0 $, with $p$ and $q$ labeling the single-particle (SP) basis. 
For the convenience of discussion, we write $H^{\dagger}$ as
\begin{equation}
	H^{\dagger} = \sum _k \langle Q_k |H | P_k \rangle b^{\dagger}_{Q_k} b_{P_k} ,
	\label{eq:labeling_k}
\end{equation}
with the mappings $Q_k \mapsto \{ u_k, v_k, \cdots , w_k \}$ and $P_k \mapsto \{p_k, q_k, \cdots , r_k \} $ such that $b^{\dagger}_{Q_k} = a ^{\dagger}_{u_k} a^{\dagger}_{v_k} \cdots a^{\dagger}_{w_k} $ and $b_{P_k} = a_{r_k} \cdots a_{q_k} a_{p_k} $.  We coordinate the indices of the monomials in Eqs. \eqref{eq:labeling_i} and \eqref{eq:labeling_k}, and set $j=k$ when $P_j = Q_k$ and $Q_j = P_k$, which provide $ (b^{\dagger}_{Q_j} b_{P_j})^{\dagger} = b^{\dagger}_{Q_k} b_{P_k}  $.

{\bf Direct encoding scheme.}
We employ the direct encoding scheme \cite{JordanWigner1928,PhysRevLett.79.2586} to map the SP basis set to a set of qubits in the quantum register. Provided a total set of $N_{\rm sp}$ SP states, each SP state corresponds to a specific qubit in the quantum register, while the vacancy and occupation of individual SP state correspond to the qubit states $\ket{0}$ and $\ket{1}$, respectively. In this sense, the Fock states are mapped to binary strings in the quantum register.

{\bf Enumerator oracle.}
We define the enumerator oracle $O_F$ as
\begin{multline}
	O_F \ket{\mathcal{F}}_s \ket{j} _{id} \ket{0}_{cp} \ket{1} _{e_p} \ket{1} _{e_q} \\ = \ket{\mathcal{F}} _s \ket{j} _{id} \ket{\mathcal{F}_j} _{ cp} \ket{y ^P_{\mathcal{F}, j}}_{e_p} \ket{y^Q_{\mathcal{F}, j}} _{e_q} ,
	\label{eq:OF_oracle}
\end{multline}
where each register is labeled by a subscript. 
With the input index $j$ and many-fermion (Fock) state $\ket{\mathcal{F}}$, $O_F$ computes the output state $\ket{\mathcal{F}_j} = b^{\dagger}_{Q_j} b_{P_j} \ket{\mathcal{F}} $. Each of ``$e_p$" and ``$e_q$" denotes a single-qubit ``error" register that is initialized as $\ket{1}$. 
If the operation $b_{P_j} \ket{\mathcal{F}} $ is valid, we flip the state of $e_p$ from $\ket{1}$ to $\ket{0}$; $e_p$ remains in $\ket{1}$ otherwise. Similarly, we flip the state of $e_q$ from $\ket{1}$ to $ \ket{0}$ if the operation $b^{\dagger}_{Q_j} \ket{\mathcal{F}_{P_j}} $ is valid; $e_q$ remains in $\ket{1}$ otherwise. As a result, $ y ^P_{\mathcal{F}, j} $ and $y^Q_{\mathcal{F}, j} $ in Eq. \eqref{eq:OF_oracle} can be either 0 or 1; it is only when the operation $b^{\dagger}_{Q_j} b_{P_j} \ket{\mathcal{F}}$ is valid that $ y ^P_{\mathcal{F}, j} = y^Q_{\mathcal{F}, j} = 0$.

\begin{figure}[ht] 
	\centering
	\includegraphics[width=\linewidth]{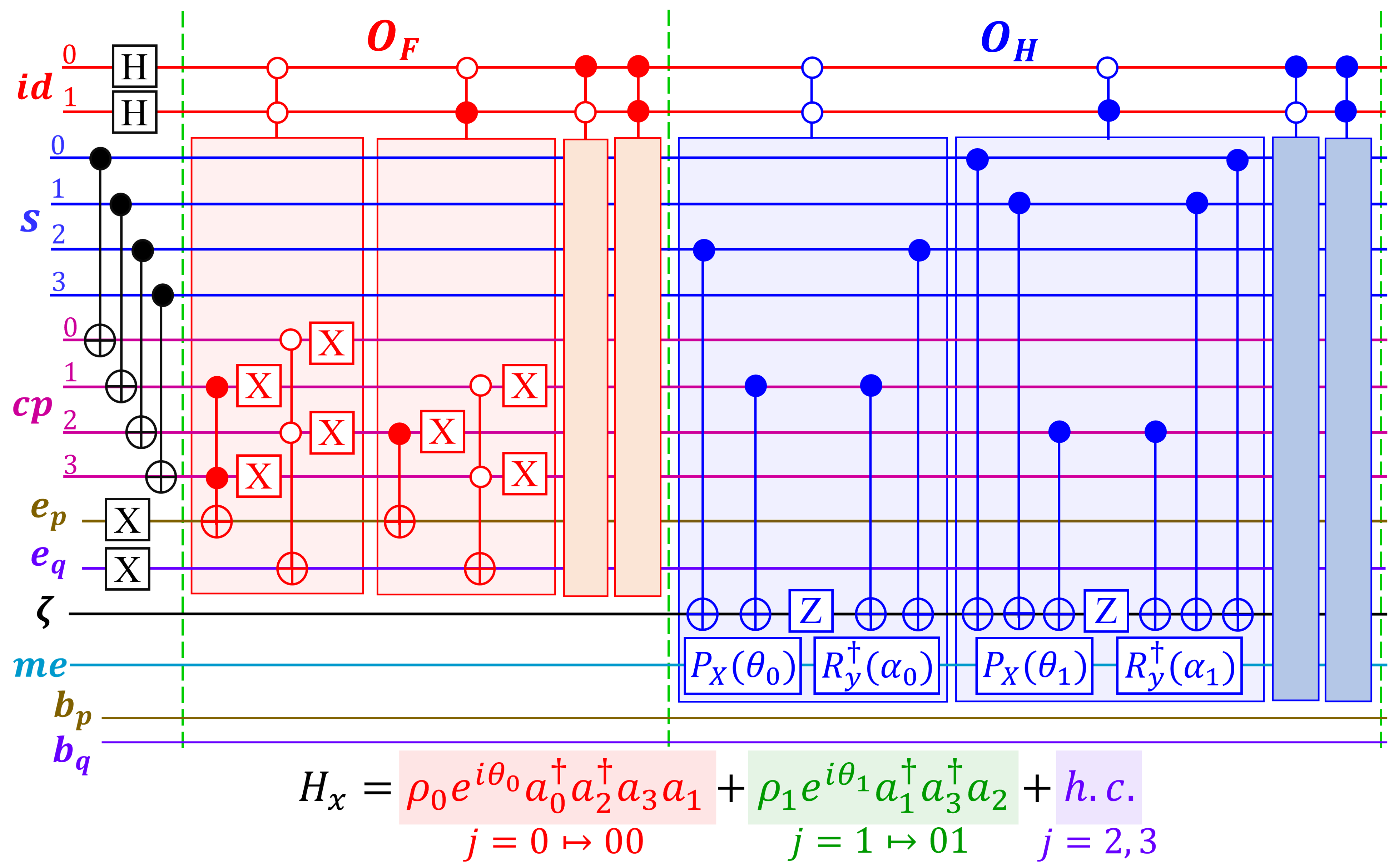}
	\caption{(color online) Illustration: the circuit of $\mathcal{T}_f$ via the $O_{\rm F}$ and $O_{ H }$ oracles designed from the Hamiltonian $H_{x}$. The qubit indices of the registers $s$ and $cp$ correspond to the SP bases. 
	} 
	\label{fig:isoforward}
\end{figure}

We construct $O_F$ as follows (see illustration in Fig. \ref{fig:isoforward}).
\paragraph*{Step 1:} Provided the Fock state $ \ket{ \mathcal{F} }$ in the ``system" register ``$s$" represented in terms of $0$'s and $1$'s, we duplicate it to the ``copy" register ``$cp$" that is initiated in the all-zero state via a sequence of qubit-wise CNOT gates. While the duplication of a general quantum state is not possible, it is possible to duplicate a simple Fock state represented in terms of $0$'s and $1$'s.

\paragraph*{Step 2:} With $j$ in the ``index" register ``$id$", we identify the $j {\rm th}$ monomial $ \langle Q_j |H | P_j \rangle b^{\dagger}_{Q_j} b_{P_j} $ in the Hamiltonian, where $P_j \mapsto \{ u_j, v_j, \cdots , w_j \}$ specifies the SP states in $ \ket{ \mathcal{F} }$ of which the occupations are to be removed.
However, one needs to check if these orbitals in $\ket{\mathcal{F}}$ are indeed occupied before the removal: if any of them is vacant, the operation $b_{P_j} \ket{\mathcal{F}}$ is invalid. We record the error message with the $e_p$ qubit that is prepared as $\ket{1}$: only if the qubits corresponds to the orbitals $\{ u_j, v_j, \cdots , w_j \}$ are all in the $\ket{1}$ states (i.e., occupied), we flip $e_p$ from $\ket{1}$ to $\ket{0}$; otherwise, $e_p$ remains in $\ket{1}$.

We have the error and validity information recorded as $y ^P_{\mathcal{F}, j} = 1 $ and $0$ by $e_p$, respectively. Then,  we proceed to flip all the qubits corresponding to the orbitals $\{ u_j, v_j, \cdots , w_j \}$ in the duplicated $\ket{\mathcal{F}}$ state in the $cp$ register by a sequence of NOT gates. The resultant state of $cp$ is $ \ket{\mathcal{F} _{P_j}} = b_{P_j} \ket{\mathcal{F}}$.

\paragraph*{Step 3:} With the index $j$, we determine a sequence of creation operators denoted by $b^{\dagger}_{Q_j}$, where $Q_j$ specifies the orbitals $\{ p_j, q_j, \cdots , r_j  \}$ in $\ket{\mathcal{F} _{P_j}}$ on which these creation operators act. 
We apply a set of multiple-open-controlled-NOT gates (i.e., those with the NOT gate controlled by a set of zero states) to check if these orbitals in $\ket{\mathcal{F} _{P_j}}$ are all vacant; that is, the qubits corresponds to these orbitals in the $cp$ register are all in the $\ket{0}$ states.
The error/validity message is recorded by the $e_q$ register which is prepared in $\ket{1}$. If the orbitals $\{ p_j, q_j, \cdots , r_j  \}$ in $\ket{\mathcal{F} _{P_j}}$ are all vacant, we flip the state of $e_q$ from $\ket{1}$ to $\ket{0}$, denoting $b^{\dagger}_{Q_j} \ket{\mathcal{F} _{P_j}}$ to be valid. Otherwise,  $e_q$ remains in $\ket{1}$, which denotes that the operation $b^{\dagger}_{Q_j} \ket{\mathcal{F} _{P_j}}$ is invalid. With the message state $ \ket{y^Q_{\mathcal{F}, j}} $ kept in $e_q$, we apply a sequence of NOT gates to flip the qubits that correspond to the orbitals  $\{ p_j, q_j, \cdots , r_j  \}$. The resultant state of $cp$ is $\ket{\mathcal{F}_j} = b^{\dagger}_{Q_j} b_{P_j} \ket{ \mathcal{F} }$.

This completes the construction of $O_F$.

{\bf Matrix element oracle.}
We define the matrix element oracle $O_H$ as
\begin{multline}
	O_H \ket{\mathcal{F}}_s \ket{j} _{id} \ket{\mathcal{F}_j} _{cp} \ket{0}_{\zeta} \ket{0}_{me} \\ =  \ket{\mathcal{F}}_s \ket{j} _{id}  \ket{\mathcal{F}_j} _{cp} { \zeta ^{{\mathcal{F}}} _{ {\mathcal{F}_j} } } \ket{0} _{\zeta} e^{ i \theta _j} \ket{\rho_{j}}_{me} ,
\end{multline}
where each register is labeled by a subscript. We have $ \zeta ^{{\mathcal{F}}} _{ {\mathcal{F}_j} } = \langle \mathcal{F}_j | b^{\dagger}_{Q_j} b_{P_j} | \mathcal{F} \rangle $ and $ \ket{\rho_{j}} \equiv {\rho } _j \ket{0} + \sqrt{1- \rho ^2 _j} \ket{1}  $. The parameters $\theta _j$ and $ \rho _j $ are determined from the few-body matrix element $ \langle Q_j |H | P_j \rangle $ of the $j$th monomial of $H$:
\begin{equation}
	h(Q_j, P_j) = \langle Q_j |H | P_j \rangle /\Lambda  = \rho _j e^{i \theta _j} ,
\end{equation}
with $\Lambda  \geq \max _j  |\langle Q_j |H | P_j \rangle | $, $ \rho _j =|h(Q_j , P_j)| \leq 1$, and $\theta _j = \arg [h(Q_j, P_j) ] \in (-\pi , \pi] $.

$ O_{ H} $ takes the monomial index $j$ in the Hamiltonian, and the Fock state $\ket{\mathcal{F}}$ and $\ket{\mathcal{F}_j}$. It computes the factor $ \zeta ^{{\mathcal{F}}} _{ {\mathcal{F}_j} } $, and encodes the matrix element as $ e^{i \theta _j} \ket{\rho _j} $. 
$ O_{ H } $ is constructed as follows (see illustration in Fig. \ref{fig:isoforward}).

\paragraph*{Step 1:} 
The factor $ \zeta ^{{\mathcal{F}}} _{ {\mathcal{F}_j} } $ is determined by two actions: $ b_{P_j} \ket{\mathcal{F}} = c(P_j, \mathcal{F}) \ket{ \bar{\mathcal{F}}'_j } $ and $ b_{Q_j} \ket{\mathcal{F}_j } = c(Q_j, \mathcal{F}_j) \ket{ \bar{\mathcal{F}}'_j }  $, where we have $ \zeta ^{{\mathcal{F}}} _{ {\mathcal{F}_j} }  = c(P_i, \mathcal{F}) \cdot c(Q_j, \mathcal{F}_j) $.

Without loss of generality, we illustrate the idea of computing $\zeta ^{{\mathcal{F}}} _{ {\mathcal{F}_j} } = \langle {\mathcal{F}_j} | b^{\dagger}_{Q_j} b_{P_j} | \mathcal{F} \rangle $ by taking $ b_{Q_j} = a_{q_j} a_{p_j}$ (with $ q_j > p_j \geq 0 $) and $b_{P_j} = a_{w_j} a_{v_j} a_{u_j} $ (with $ w_j >v_j > u_j \geq 0$). For the case $ a_{w_j} a_{v_j} a_{u_j} \ket{\mathcal{F}} $, we group the fermion operators in pairs from left to right, and obtain $ ( a_{w_j} a_{v_j} ) a_{u_j} $ with $ a_{u_j} $ being unpaired. Then, in the Fock state $\ket{\mathcal{F}}$, we count the occupancy $N^{\rm occ}_{0\rightarrow u_j}$ in all the orbitals with the indices $\kappa \in (0, u_j)$, and the occupancy $N^{\rm occ}_{v_j \rightarrow w_j}$ in all the orbitals with the indices $ \kappa  \in (v_j,w_j) $. The same technique applies to the other case $ a_{q_j} a_{p_j} \ket{\mathcal{F}_j}$. We first group the string of the fermion operators in pairs from the left to right, where no operator is left unpaired in this case. We then count the occupancy $N^{\rm occ}_{p_j \rightarrow q_j}$ in all the orbitals with the indices $\eta \in (p_j,q_j)$ in $ \ket{\mathcal{F}_j}$. 

With that, we record $ N^{\rm occ}_{{\rm tot},j} = N^{\rm occ}_{0\rightarrow u_j} + N^{\rm occ}_{v_j \rightarrow w_j} + N^{\rm occ}_{p_j \rightarrow q_j} $ as the state $ \ket{ {\rm mod }(N^{\rm occ}_{{\rm tot},j},2) } $ by the single-qubit register $\zeta $ that is initialized as $\ket{0}$. 
This is achieved by two sequences of CNOT gates: (1) in one set, each CNOT is controlled by the $\kappa $th qubit in the $s$ register that encodes $\ket{\mathcal{F}}$, and acts on the single-qubit register $\zeta $, where $ \kappa \in (0, u_j) \cup (v_j,w_j) $; and (2) in the other set, each CNOT is controlled by the $\eta $th qubit in the $cp$ register that encodes $\ket{\mathcal{F}_j}$, and acts on $\zeta $, where $\eta \in (p_j,q_j)$.

The factor $\zeta ^{{\mathcal{F}}} _{ {\mathcal{F}_j} } =  (-1)^ {{\rm mod }(N^{\rm occ}_{{\rm tot},j},2) } $ is obtained by applying a Pauli-$Z$ gate on the state $\ket{ {\rm mod }(N^{\rm occ}_{{\rm tot},j},2) } $ of $\zeta $. Finally, we uncompute the state of $\zeta $ to $\ket{0}$.

The calculations of $\zeta ^{{\mathcal{F}}} _{ {\mathcal{F}_j} } $ in general cases follow the technique discussed above. 

\paragraph*{Step 2.}
The matrix element $\rho _j e^{i \theta _j}$ of the $j$th monomial is recorded as the coefficient of the $\ket{0}$ component of the state $ e^{i\theta _j} \ket{ \rho_j }$ by the single-qubit ``matrix-element" register ``$me$". This is achieved by applying a phase gate $P_X (\theta _j)$ and then a rotation-Y gate $R^{\dagger}_y ( \alpha _j )$ \cite{Qiskit} to the register $me$ that is initialized in $\ket{0}$, where we define $P_X (\theta _j) \ket{0} =  e^{i\theta _j} \ket{0} $ and $P_X (\theta _j) \ket{1} = \ket{1}$, and we take $\alpha _j = 2 \arccos {\rho_{j}}$. 

This completes the construction of the $O_H$ oracle.

{\bf Walk-state constructions.}
With the oracles, we introduce the isometries to construct the quantum walk states. We define the isometry $\mathcal{T}_f$ to create the ``forward" walk state with the Fock state $ | \mathcal{F} \rangle $ as
\begin{align}
	\mathcal{T}_f | \mathcal{F} \rangle _s \otimes  & \ket{0}_a =  \frac{1}{\sqrt{\mathcal{D}}} \sum _{j=0}^{\mathcal{D}-1} { \zeta ^{{\mathcal{F}}} _{ {\mathcal{F}_j} }  } e^{i \theta _{j}} \ket{\mathcal{F}}_s  \ket{j}_{id} \ket{\mathcal{F}_{j}}_{cp}  \nonumber \\
	& \otimes \ket{y ^P_{\mathcal{F}, j}}_{e_p}  \ket{y ^Q_{\mathcal{F}, j}}_{e_q}  \ket{0}_{\zeta } \ket{\rho_{j}} _{me} \ket{0}_{b_p} \ket{0}_{b_q} ,
	\label{eq:forwarddd_state}
\end{align}
with one query to both the $O_F$ and $O_H$ oracles constructed based on $H$. Except for the register $s$, the ancilla register (denoted by ``$a$") consists of the other eight registers. Here, the ``$b_p$" and ``$b_q$" registers are both prepared in $\ket{0}$. These registers are employed to remove the non-physical walk-state components labeled by $\ket{1}_{e_p} $ and  $\ket{1}_{e_q} $.  $\mathcal{D} $ denotes the total number of monomials in $H$.

Similarly, we define the isometry $\mathcal{T}_b $ to create the ``backward" walk state with the Fock state $ | \mathcal{G} \rangle $ as
\begin{align}
	\mathcal{T}_b | \mathcal{G} \rangle _s \otimes \ket{0}_a  = \frac{1}{\sqrt{\mathcal{D}}} & \sum _{k=0}^{\mathcal{D}-1} \ket{\mathcal{G}} _s \ket{k}_{id} \ket{\mathcal{G}_{k}} _{cp} \ket{ y ^P_{\mathcal{G}, k}  } _{e_p} \nonumber \\ 
	& \otimes \ket{ y ^Q_{\mathcal{G}, k} } _{e_q} \ket{0} _{\zeta } \ket{0} _{me } \ket{0} _{b_p} \ket{0} _{b_q} 	,
	\label{eq:backwarddd_state}
\end{align}
with one query to the $O_F$ oracle that is constructed based on $H^{\dagger}$. We recall that the monomial labeling schemes in $H$ and $H^{\dagger}$ are coordinated. 
More details of our construction of the walk states are available in the Appendix \ref{sec:walkstate_constructions_appendix}.

{\bf Hamiltonian input scheme.}
These quantum-walk states [Eqs. \eqref{eq:forwarddd_state} and \eqref{eq:backwarddd_state}] block encode $H$ as (see the proof in Appendix \ref{sec:block_encoding_via_Walk_States})
\begin{equation}
	( \langle \mathcal{G} |_s \otimes \langle 0| _a ) \mathcal{T}^{\dagger}_b {S} \mathcal{T}_f ( | \mathcal{F} \rangle _s \otimes \ket{0} _a ) = \frac{1}{\mathcal{D} \Lambda } \langle  \mathcal{G} | H | \mathcal{F} \rangle ,
\end{equation}
with $ {1}/{(\mathcal{D} \Lambda )} $ being the scaling coefficient.
This defines our walk-state-based Hamiltonian input scheme $U_H \equiv \mathcal{T}^{\dagger}_b {S} \mathcal{T}_f $ with $U_H \neq U^{\dagger}_H$. $S$ denotes a set of bit-wise swap operations: (1) between the $s$ and $cp$ registers; (2) between the single-qubit registers $e_p $ and $b_p$; and (3) between single-qubit registers $e_q $ and $b_q$. $S$ prunes the walk-state components according to the symmetries of $H$ as well as the error messages encoded by the registers $e_p$ and $e_q$, such that only those physical walk-state components are retained.

{\bf Spectral calculations.}
We propose the SA-QKSD framework for the spectral calculations of relativistic and nonrelativistic many-fermion systems with the synergy between our input scheme and the QKSD method \cite{Kirby_2023}. Readers are also referred to Appendix \ref{sec:SA_QKSD_framework} for more details.

Based on our input scheme and the qubitization \cite{Low2019hamiltonian,gilyen2019quantum,PRXQuantum.2.040203}, we can block encode the Chebyshev polynomials of the first kind $T_i(H')$ as a function of $H'= H/(\mathcal{D}\Lambda)$. Then, we can construct the Krylov bases based on the pivot Fock state $ \ket{\psi _0} $ on the quantum computer as
\begin{equation}
	\mathcal{B}_{\mathcal{K}} = \{ \ket{\psi _i} = T_i (H') \ket{\psi _0} | i = 0, 1, 2,  \cdots , \mathcal{K} -1  \} 
\end{equation}
We note that our input scheme of $T_i (H')$ respects the symmetries of the Hamiltonian. It is also straightforward to manipulate the symmetries of $\ket{\psi _0}$ in the occupation representation. These enable direct control on the symmetries of $\mathcal{B}_{\mathcal{K}}$ and the Hilbert subspace spanned by $\mathcal{B}_{\mathcal{K}}$. 

In the SA Krylov basis, we can evaluate the expectations $ \langle T_k  (H') \rangle _0 \equiv \langle \psi _0 | T_k (H') | \psi _0 \rangle $ with the Hadamard test \cite{nielsen2010quantum}  on the quantum computer. This handles the many-body calculations that are challenging for classical computers. These expectations $\{ \langle T_k  (H') \rangle _0 \}$ are input to the classical computer to calculate the Hamiltonian and overlap matrix elements, based on which one constructs the generalized eigenvalue equation with greatly reduced dimension for the lowest eigenvalue of $H'$ \cite{doi:10.1137/21M145954X} in the SA Hilbert subspace. 

As an example, $H$ preserves the particle number and the projection $M_J$ of the total angular momentum $J$ of the many-nucleon systems in nuclear structure calculations. We can generate the SA Krylov bases with specific $M_J$ by electing $\ket{\psi _0}$ to be of fixed $M_J$. 
With this $M_J$ regulation, those eigenstates of $H$ with $J < M_J$ are excluded in the resultant spectrum. This SA screening technique is useful to probe the excited states. 
Additional options for symmetry regulations can be found in, e.g., Refs. \cite{Lipkin:1958,Gloeckner:1974sst,Wang:2024scd}.

{\bf Gate complexity analysis.}
We analyze the gate complexity of our input scheme $U_H$ and $U^{\dagger}_H$. We consider the extreme case where all the $\mathcal{D}$ monomials in $ H $ are terms with $\mathcal{X}$ creation and $\mathcal{Y}$ annihilation operators, together with their conjugate terms. The gate cost to construct $O_F$ and $O_H$ are $\widetilde{O}\big( \mathcal{D} \big)$ and $\widetilde{O}\big( \mathcal{D} \cdot N_{\rm sp} \big)$, respectively. Here we have $ \mathcal{D} \in O(N^{\mathcal{X}+ \mathcal{Y} }_{\rm sp}) $ and the $\widetilde{O}$ notation means that the logarithmic factor from compiling the multiple controlled gates \cite{10.1145/2597917.2597939,javadiabhari2015scaffcc} is suppressed. Therefore, the gate cost to construct $U_H$ and $U^{\dagger}_{H}$ is $\widetilde{O}\big( N_{\rm sp}^{\mathcal{X}+ \mathcal{Y}+1} \big)$. Indeed, our input scheme provides a rigorous bound of the gate cost for encoding arbitrary second-quantized many-fermion Hamiltonians such as those in field theory problems, where many-body forces and particle-number variations may be present. 

For inputting the Hamiltonian which includes two-fermion interactions at most (e.g., the molecular Hamiltonian) in an arbitrary basis, the lowest rigorous gate complexity from prior works is $\widetilde{O}(N_{\rm sp}^5)$ \cite{Babbush_2016,Berry_2019}. This complexity is achieved by the linear combination of unitaries (LCU) \cite{childs2012hamiltonian} employing either the Jordan–Wigner transformation \cite{JordanWigner1928, PhysRevLett.79.2586} or the Bravyi–Kitaev transformation \cite{Bravyi_2002,BK_encodingScheme}. Our Hamiltonian input scheme reports the same gate cost (with $\mathcal{X}=\mathcal{Y}=2$). 

We expect that our input scheme is simpler than the standard approaches utilizing the Jordan–Wigner or Bravyi–Kitaev transformation, especially when many-body interactions (e.g., in quantum field theory and nuclear physics applications) are present. This simplification results from the fact that our scheme encodes the monomials and the few-body matrix elements in a straightforward manner, where each monomial in the Hamiltonian maps to a set of controlled gates for qubit-state tests, qubit flipping, phase operations, and rotations. Unlike the Jordan–Wigner or Bravyi–Kitaev transformation, our scheme does not decompose the monomial of fermion operators into multiple Pauli strings.
Due to this difference in encoding the Hamiltonian monomials, our oracle designs differ from the ``prepare" and ``select" oracles in the standard LCU; we expect that our oracle designs can provide new insights into the circuit construction of the LCU.

The gate complexity of our input scheme can be further reduced by adopting the symmetry properties of Hamiltonians, by employing SP basis tailored to the Hamiltonian symmetries, and by implementing improved basis-qubit mapping schemes \cite{Babbush_2016,Berry_2019, PhysRevLett.123.070503,PhysRevX.8.011044}. Meanwhile, advanced oracle design strategies [cf., Ref. \cite{Wan_2021}] could also be helpful in reducing the gate complexity of our input scheme. These are open research fields for future advancement.

As for our framework of spectral calculations, we evaluate $ \langle T_k  (H') \rangle _0$ with $ k = 0, 1, 2, \cdots , 2\mathcal{K}-1$ on the quantum computer. A general bound on $\mathcal{K}$ to compute the ground state energy within the error $\mathcal{E}$ is \cite{Kirby_2023}
\begin{equation}
	\mathcal{K} = \Theta [  ( \log | \gamma _0 |^{-1}  + \log {\mathcal{E}}^{-1}  ) \cdot \min ( {\mathcal{E}}^{-1}, {\Delta}^{-1} )  ] ,
	\label{eq:dim_or_KrylovBasisSpace}
\end{equation}
with $\Delta $ being the spectral gap. $ \gamma_0 $ denotes the overlap between the pivot $\ket{\psi_0 }$ and the ground state of $H'$ in the SA Hilbert subspace. 

The total number of the quantum evaluations of $ \langle T_k  (H') \rangle _0 $ scales as $\Theta ( \mathcal{K}^2 )$. The gate complexity of the most complex evaluation scales as $ \widetilde{O} \big( \mathcal{K} \cdot N^{\mathcal{X}+ \mathcal{Y} +1 }_{\rm sp}   \big) $.

{\bf Nuclear structure calculations.}
We demonstrate our framework by solving the low-lying spectra of $^{42}$Ca and $^{46}$Ca. These isotopes are considered as the two- and six-neutron systems outside of a closed ${^{40}}$Ca core \cite{suhonen2007nucleons}. 
While we restrict our demonstrations to these two limited examples, our hybrid framework is applicable for general first-principles nuclear structure calculations utilizing inter-nucleon interactions from fundamental theories on future fault-tolerant quantum hardware.

We model the Hamiltonian as the pairing plus quadrupole-quadrupole interaction \cite{Dreiss:1971xe}
\begin{align}
	&	H_A =  g \Big[ - \sum _{p , q } \xi _{p } \xi _{q} a_{p}^{\dagger} a_{\bar{p}}^{\dagger} a _{\bar{q}} a_q \nonumber \\
	& + \chi \sum _{p<q} \sum _{u < v} \sum _{\mu =-2}^2 \langle p  | \mathcal{Q}_{2\mu} | u \rangle  \langle q | \mathcal{Q}^{\ast}_{2\mu} | v \rangle a^{\dagger}_{p} a^{\dagger}_{q} a_{v} a_{u } \Big] ,
	\label{eq:HA}
\end{align}
with the quadrupole operator $ \mathcal{Q}_{2\mu} \equiv r^2 Y_{2\mu } (\Omega _{\hat{r}}) $.
We take the $0f_{7/2}$ valence space for the calculation and adopt the harmonic oscillator basis \cite{suhonen2007nucleons,Du:2018tce} as the SP basis. $a_{p}$ annihilates a neutron on the $ p $th SP basis labeled by $ p \mapsto (n_{p}, l_{p}, j_{p}, m_{p})$ with the radial quantum number $n_p$, the orbital angular momentum $l_p$, the total angular momentum $j_p$ and its projection $m_p$ (with $m_p>0$), where the quantum numbers of the spin and isospin are omitted. We also define the label $\bar{p } \mapsto (n_{p}, l_{p}, j_{p}, - m_{p}) $, and the coefficient $\xi _{p } \equiv (-1)^{j_{p} - m_{p}}$. We take $ g = 0.147439 $ MeV and $ \chi = -3.934 \times 10^8 $ by fitting the excitation spectrum of $^{42}$Ca. The same $g$ and $\chi$ values are adopted to compute the spectra of $^{46}$Ca.

We elect the oscillator strength and nucleon mass to be 12 MeV and 938.919 MeV, respectively. We evaluate the two-body matrix elements of $H_A$ with the retained SP basis set in the $0f_{7/2}$ valence shell (Table \ref{tab:SPbasis_0f} in Appendix \ref{sec:numerical_calculations_info}), where these two-body matrix elements are listed in Table \ref{tab:ME_table} in Appendix \ref{sec:numerical_calculations_info}.
We construct the circuits to block encode the Chebyshev polynomials of the target Hamiltonian based on our input scheme.
By selecting the symmetries of the pivot state $\ket{\psi _0}$, we construct the SA Krylov basis sets, with which we evaluate the Hamiltonian and overlap matrix elements utilizing the noiseless Statevector simulator in the IBM Qiskit \cite{Qiskit}. These matrix elements are input to the classical computer to solve the eigenenergies. 

\begin{table}[ht]
	\centering
	\caption{The low-lying eigenenergies and excitation spectra (in MeV's) of $^{42}$Ca and $^{46}$Ca compared to the experiment \cite{NNDC2022}.}
	\begin{tabular}{c|ccc|ccc}
		\hline \hline
		$J^{\pi}$ & $E_{^{42}Ca}$         & $E^{\rm ex}_{^{42}Ca}$   & Expt.   & $E_{^{46}Ca}$        & $E^{\rm ex}_{^{46}Ca}$  & Expt.  \\ \hline
		$0^+$     & {\bf -2.34280}  & 0       & 0       & {\bf 0.868409} & 0       & 0      \\
		$2^+$     & {\bf-0.818086}  & 1.52471 & 1.52471 & {\bf 2.39312}  & 1.52471 & 1.346  \\
		$4^+$     & {\bf 0.584347}  & 2.92714 & 2.7524  & {\bf 3.79555}  & 2.92714 & 2.5747 \\
		$6^+$     & {\bf 0.584347}  & 2.92714 & 3.18926 & {\bf 3.79555}  & 2.92714 & 2.9739 \\
		\hline \hline
	\end{tabular}
	\label{tab:spectra_table_2n6n}
\end{table}

We obtain the eigenenergies of the $^{42}$Ca and $^{46}$Ca with our hybrid framework.
We find the results via our hybrid framework (bold font in Table \ref{tab:spectra_table_2n6n}) agree with the classical calculations via the full configuration interaction method to 6 significant figures (classical calculation results not shown in Table I); they also compare well with the experiment \cite{NNDC2022}. We note the excitation spectra of the two isotopes are identical. This is expected from the fact that, in the $0f_{7/2}$ valence space, the two- and six-neutron systems are conjugate two-particle and two-hole systems, respectively.

{\bf Summary and outlook.}
In sum, we propose a novel input scheme for general second-quantized Hamiltonians of relativistic or non-relativistic many-fermion systems. Working in the occupation representation, our input scheme respects the symmetries of Hamiltonians, and incorporates the Pauli principle, anticommutation relations, and particle number variations. 
Based on our input scheme, we propose a hybrid framework for the spectral solutions of the many-fermion systems in the relativistic and nonrelativistic quantum theories on future quantum hardwares.
We present explicit circuit designs of our input scheme and hybrid framework of spectral calculations, where the gate complexity is analyzed.
We showcase our hybrid framework by computing the low-lying spectra of $^{42}$Ca and $^{46}$Ca. 
Our input scheme provides new pathways to solving the time-evolution unitary \cite{PhysRevLett.118.010501,gilyen2019quantum,Du:2023bpw} of the relativistic and nonrelativistic many-fermion systems whose Hamiltonians and basis states are best described in second quantization for the structure and dynamics via efficient quantum algorithms.

{\bf Acknowledgments.}
We acknowledge fruitful discussions with Peter Love, Chao Yang, Pieter Maris, Michael Kreshchuk, and William Kirby.
This work was supported by US DOE Grant DE-SC0023707 under the Office of Nuclear Physics Quantum Horizons program for the ``{\bf Nu}clei and {\bf Ha}drons with {\bf Q}uantum computers ({\bf NuHaQ})" project.


\appendix

\section{Constructions of the quantum walk states}
\label{sec:walkstate_constructions_appendix}

We provide the proof of Eqs. (6) and (7) appearing in the main text:
\begin{align}
	\mathcal{T}_f | \mathcal{F} \rangle _s \otimes   \ket{0}_a =&  \frac{1}{\sqrt{\mathcal{D}}} \sum _{j=0}^{\mathcal{D}-1} { \zeta ^{{\mathcal{F}}} _{ {\mathcal{F}_j} }  } e^{i \theta _{j}} \ket{\mathcal{F}}_s  \ket{j}_{id} \ket{\mathcal{F}_{j}}_{cp}   \otimes \ket{y ^P_{\mathcal{F}, j}}_{e_p}  \ket{y ^Q_{\mathcal{F}, j}}_{e_q}  \ket{0}_{\zeta } \ket{\rho_{j}} _{me} \ket{0}_{b_p} \ket{0}_{b_q} , \label{eq:fowardWalkState}
	\\
	\mathcal{T}_b | \mathcal{G} \rangle _s \otimes \ket{0}_a  =& \frac{1}{\sqrt{\mathcal{D}}}  \sum _{k=0}^{\mathcal{D}-1} \ket{\mathcal{G}} _s \ket{k}_{id} \ket{\mathcal{G}_{k}} _{cp} \ket{ y ^P_{\mathcal{G}, k}  } _{e_p} 
	\otimes \ket{ y ^Q_{\mathcal{G}, k} } _{e_q} \ket{0} _{\zeta } \ket{0} _{me } \ket{0} _{b_p} \ket{0} _{b_q} . \label{eq:backwardWalkState}
\end{align}
We prove these equations by straightforward constructions.

\begin{proof}
	
	In the following, we first construct the walk states based on $\ket{\mathcal{F}}$ and $\ket{\mathcal{G}}$, where the walk states can be written in terms of the isometries $\mathcal{T}_f $ and $\mathcal{T}_b $.
	
	\paragraph*{Case 1: Preparation of the forward walk state [Eq. \eqref{eq:fowardWalkState}].} 
	
	\begin{enumerate}
		\item 
		Provided the many-nucleon state $\ket{\mathcal{F}} $, the input state of the quantum computer to be
		\begin{equation}
			\ket{\Psi _1} = | \mathcal{F} \rangle _s \otimes \ket{0}_a = \ket{0} _{id} \ket{\mathcal{F}}_s \ket{0} _{cp} \ket{0} _{e_p} \ket{0} _{e_q} \ket{0} _{\zeta } \ket{0} _{me} \ket{0} _{b_p} \ket{0} _{b_q} , 
			\label{eq:step1_forward}
		\end{equation} 
		where we specify the initial state of each quantum register explicitly. The ancilla register $a$ contains the registers $id$, $cp$, $e_p$, $e_q$, $\zeta$, $me$, $b_p$, and $b_q$.
		
		\item 
		We flip the $4^{\rm th}$ and $5^{\rm th}$ registers to the states $\ket{1}$ by applying two Pauli-X gates. 
		\begin{equation}
			\ket{\Psi _{2}} = \ket{0} _{id} \ket{\mathcal{F}} _s \ket{0} _{cp} \ket{1}_{e_p} \ket{1} _{e_q} \ket{0} _{\zeta } \ket{0} _{me} \ket{0} _{b_p} \ket{0} _{b_q}  . 
			\label{eq:step2_forward}
		\end{equation}
		
		\item 
		We prepare a uniform superposition of the indices $j = 0,1,2, \cdots , \mathcal{D}-1$ by applying the diffusion operator (a sequence of Hadamard gates) to the register $id$. The overall state becomes
		\begin{equation}
			\ket{\Psi _{3,f}} = \frac{1}{\sqrt{\mathcal{D}}} \sum _{j=0}^{\mathcal{D}-1} \ket{ j }_{id} \ket{\mathcal{F}} _s \ket{0} _{cp} \ket{1}_{e_p} \ket{1}_{e_q} \ket{0}_{\zeta} \ket{0}_{me} \ket{0}_{b_p} \ket{0} _{b_q} .
		\end{equation}
		
		\item 
		We apply the $O_{F}$ oracle [Eq. (3) in the main text] to the first five registers in the summation to have the overall state of the registers as
		\begin{equation}
			\ket{\Psi _{4,f}} = \frac{1}{\sqrt{\mathcal{D}}} \sum _{j = 0}^{\mathcal{D}-1} \ket{j}_{id} \ket{\mathcal{F}}_s \ket{\mathcal{F}_{j}}_{cp} \ket{y_P(\mathcal{F},j)}_{e_p} \ket{y_Q(\mathcal{F},j)}_{e_q} \ket{0}_{\zeta} \ket{0}_{me} \ket{0}_{b_p} \ket{0}_{b_q} ,
		\end{equation}
		with $y_P(\mathcal{F},j) =0 $ or $1$, and $y_Q(\mathcal{F},j) =0 $ or $1$. The oracle $O_F$ is constructed based on $H$ [Eq. (1) in the main text].
		
		\item 
		We apply the $O_H$ oracle to the registers $id$, $s$, $cp$, $\zeta $, and $me$ according to Eq. (4) in the main text. The overall state of the registers becomes
		\begin{equation}
			\ket{\Psi _{5,f}} = \frac{1}{\sqrt{\mathcal{D}}} \sum _{j=0}^{\mathcal{D}-1} \ket{j}_{id} \ket{\mathcal{F}}_s \ket{\mathcal{F}_{j}}_{cp} \ket{y_P(\mathcal{F},j)}_{e_p} \ket{y_Q(\mathcal{F},j)}_{e_q} { \ \zeta _{\mathcal{F}_j}^{\mathcal{F}}  \ } \ket{0}_{\zeta} e^{i \theta _{j}} \big[ {\rho _{j} } \ket{0}_{me} + \sqrt{1- \rho ^2_{j}} \ket{1}_{me}  \big] \ket{0}_{b_p} \ket{0}_{b_q} .
			\label{eq:step5_forward}
		\end{equation}

	\end{enumerate}

	\paragraph*{Case 2: Preparation of the backward walk state [Eq. \eqref{eq:backwardWalkState}].}
	
	\begin{enumerate}
		
		\item 
		Provided the many-nucleon state $\ket{\mathcal{G}}$, we initialize the input state of the quantum computer to be
		\begin{equation}
			\ket{\Phi _1} = | \mathcal{G} \rangle _s \otimes \ket{0}_a = \ket{0} _{id} \ket{\mathcal{G}}_s \ket{0} _{cp} \ket{0} _{e_p} \ket{0} _{e_q} \ket{0} _{\zeta} \ket{0} _{me} \ket{0} _{b_p} \ket{0} _{b_q} , 
		\end{equation} 
		where we specify the initial state of each quantum register explicitly.	
		
		\item 
		We flip the $4^{\rm th}$ and $5^{\rm th}$ registers to the states $\ket{1}$ by applying two Pauli-X gates. 
		\begin{equation}
			\ket{\Phi _{2}} = \ket{0} _{id} \ket{\mathcal{G}} _s \ket{0} _{cp} \ket{1} _{e_p} \ket{1} _{e_q} \ket{0} _{\zeta} \ket{0} _{me} \ket{0} _{b_p} \ket{0} _{b_q} . \label{eq:step2}
		\end{equation}
		
		\item 	
		We prepare a uniform superposition of the indices $k=0,1,2, \cdots , \mathcal{D}-1$ by applying the diffusion operator on the register $id$. The overall state becomes
		\begin{equation}
			\ket{\Phi _{3,b}} = \frac{1}{\sqrt{\mathcal{D}}} \sum _{k=0}^{\mathcal{D}-1} \ket{k}_{id} \ket{\mathcal{G}} _s \ket{0} _{cp} \ket{1} _{e_p} \ket{1} _{e_q} \ket{0}_{\zeta} \ket{0} _{me} \ket{0} _{b_p} \ket{0} _{b_q}.
		\end{equation}
		
		\item 
		We apply the $O_{F}$ oracle to the first five registers to have the overall state of the registers as
		\begin{equation}
			\ket{\Phi _{4,b}} = \frac{1}{\sqrt{\mathcal{D}}} \sum _{k=0}^{\mathcal{D}-1} \ket{k}_{id} \ket{\mathcal{G}} _s \ket{\mathcal{G}_{k}} _{cp} \ket{y_P(\mathcal{G},k)}_{e_p} \ket{y_Q(\mathcal{G},k)} _{e_q} \ket{0}_{\zeta} \ket{0}_{me} \ket{0}_{b_p} \ket{0}_{b_q} ,
			\label{eq:step4_backward}
		\end{equation}
		with $y_P(\mathcal{G},k) =0 $ or $1$, and $y_Q(\mathcal{G},k) =0 $ or $1$. Note that here the oracle $O_F$ is constructed based on $H^{\dagger}$ [Eq. (2) in the main text].

	\end{enumerate}
	
	Equations \eqref{eq:step5_forward} and \eqref{eq:step4_backward} are Eqs. \eqref{eq:fowardWalkState} and \eqref{eq:backwardWalkState}, respectively. This completes the proof.
	
\end{proof}

We can extract $| \mathcal{F} \rangle $ in $\ket{\Psi _{5,f}} $ and rewrite $ \ket{\Psi _{5,f}} $ in terms of the isometry $\mathcal{T}_f $ as
\begin{equation}
	\mathcal{T}_f | \mathcal{F} \rangle _s \otimes \ket{0}_a = | \mathcal{F} \rangle _s \otimes \ket{\phi _{\mathcal{F},f}} _a , 
	\label{eq:isometryT_f}
\end{equation}
where $\ket{\phi _{\mathcal{F},f}} _a $ absorbs the terms other than  $| \mathcal{F} \rangle $ on the right hand side of Eq. \eqref{eq:step5_forward}. 

Similarly, after extracting $\ket{\mathcal{G}}$ in $\ket{\Phi _{4,b}} $, we can also rewrite $\ket{\Phi _{4,b}} $ with the isometry $\mathcal{T}_b$ as
\begin{equation}
	\mathcal{T}_b | \mathcal{G} \rangle _s \otimes \ket{0}_a = | \mathcal{G} \rangle _s \otimes \ket{\phi _{\mathcal{G},b}} _a ,
	\label{eq:isometryT_b}
\end{equation} 
where $ \ket{\phi _{\mathcal{G},b}} $ absorbs all the terms (except $\ket{\mathcal{G}}$) on the right hand side of Eq. \eqref{eq:step4_backward}.

\section{Hamiltonian input scheme}
\label{sec:block_encoding_via_Walk_States}

Our Hamiltonian input scheme for the many-fermion system can be written as
\begin{equation}
	( \langle \mathcal{G} |_s \otimes \langle 0| _a ) \big( \mathcal{T}^{\dagger}_b {S} \mathcal{T}_f \big) ( | \mathcal{F} \rangle _s \otimes \ket{0} _a ) = \frac{1}{\mathcal{D} \Lambda _m} \langle  \mathcal{G} | H | \mathcal{F} \rangle ,
	\label{eq:BE_Hamiltonian}
\end{equation}
where $S$ denotes the swap operator, of which the definition is shown in the following text. 
We note that $U_H \equiv \mathcal{T}^{\dagger}_b {S} \mathcal{T}_f $ block encodes the Hamiltonian $H ' = H/(\mathcal{D} \Lambda _m) $. It is also worth noting that $U_H^{\dagger} \neq U_H$.

\begin{proof}
	We prove that the walk states $ \mathcal{T}_f | \mathcal{F} \rangle _s \otimes \ket{0} _a = \ket{\Psi _{5,f}}  $ and $ \mathcal{T}_b | \mathcal{G} \rangle _s \otimes \ket{0} _a = \ket{\Phi _{4,b}}$ can be applied to block encode the input Hamiltonian according to Eq. \eqref{eq:BE_Hamiltonian}, with proper definition of the swap operator ${S}$. This can be achieved by straightforward calculations. We have from the left hand side of Eq. \eqref{eq:BE_Hamiltonian} 
	\begin{align}
		&	( \langle \mathcal{G} |_s \otimes \langle 0| _a ) \big( \mathcal{T}^{\dagger}_b {S} \mathcal{T}_f \big) ( | \mathcal{F} \rangle _s \otimes \ket{0} _a ) \nonumber \\
		=& \langle \Phi _{4,b} | S | \Psi_{5,f} \rangle \nonumber \\
		=&  \left\{  \frac{1}{\sqrt{\mathcal{D}}} \sum _{k=0}^{\mathcal{D}-1} \ket{k}_{id} \ket{\mathcal{G}} _s \ket{\mathcal{G}_{k}}_{cp} \ket{y_P(\mathcal{G},k)}_{e_p} \ket{y_Q(\mathcal{G},k)} _{e_q} \ket{0}_{\zeta} \ket{0}_{me} \ket{0} _{b_p} \ket{0}_{b_q}  \right\} ^{\dagger} \nonumber \\
		& \mathcal{S} \left\{ \frac{1}{\sqrt{\mathcal{D}}} \sum _{j=0}^{\mathcal{D}-1} \ket{j}_{id} \ket{\mathcal{F}}_s \ket{\mathcal{F}_{j}} _{cp} \ket{y_P(\mathcal{F},j)}_{e_p} \ket{y_Q(\mathcal{F},j)} _{e_q} { \ \zeta _{\mathcal{F}_j}^{\mathcal{F}} \ } \ket{0}_{\zeta} e^{i \theta _{j}} \big[ {\rho _{j} } \ket{0}_{me} + \sqrt{1- \rho ^2 _{j}} \ket{1}_{me}  \big] \ket{0}_{b_p} \ket{0}_{b_q}   \right\} .
		\label{eq:mid20}
	\end{align}
	We define the swap operator $ {S} $ as
	\begin{equation}
		{S} \ket{r_1}_{0} \ket{r_2}_{1} \ket{r_3}_{2} \ket{r_4}_{3} \ket{r_5}_{4} \ket{r_6}_{5} \ket{r_7}_6 \ket{r_8}_7  \ket{r_{9}}_8 = \ket{ r_1}_0 \ket{r_3}_1 \ket{r_2}_2 \ket{r_8}_3 \ket{r_{9}}_4 \ket{r_6}_5  \ket{r_7}_6 \ket{r_4}_7 \ket{r_{5}}_8 .
		\label{eq:swap_operation}
	\end{equation}
	
	Then, we have
	\begin{align}
		{S} \Bigg\{ 
		&  \frac{1}{\sqrt{\mathcal{D}}} \sum _{j=0}^{\mathcal{D}-1} e^{i\theta _{j}}  \sqrt{1- \rho ^2_{j}}  { \ \zeta _{\mathcal{F}_j}^{\mathcal{F}} \ } \ket{j}_{id} \ket{\mathcal{F}}_s \ket{\mathcal{F}_{j}}_{cp} \ket{y_P(\mathcal{F},j)}_{e_p} \ket{y_Q(\mathcal{F},j)}_{e_q} \ket{0} _{\zeta}  \ket{1} _{me} \ket{0} _{b_p} \ket{0}_{b_q} 
		\nonumber \\ 
		+& \frac{1}{\sqrt{\mathcal{D}}} \sum _{j=0}^{\mathcal{D}-1}  e^{i\theta _{j}} {\zeta _{\mathcal{F}_j}^{\mathcal{F}}  } { \rho _{j} } \ket{j}_{id} \ket{\mathcal{F}}_s \ket{\mathcal{F}_{j}} _{cp} \ket{y_P(\mathcal{F},j)}_{e_p} \ket{y_Q(\mathcal{F},j)}_{e_q} \ket{0}_{\zeta}  \ket{0}_{me}  \ket{0}_{b_p} \ket{0}_{b_q} 
		\Bigg\} \nonumber \\
		= & \frac{1}{\sqrt{\mathcal{D}}} \sum _{j=0}^{\mathcal{D}-1} e^{i\theta _{j}}  \sqrt{1- \rho ^2 _{j}} \ { \zeta _{\mathcal{F}_j}^{\mathcal{F}}  } \ket{ j }_{id}  \ket{\mathcal{F}_{j}}_s \ket{\mathcal{F}}_{cp}  \ket{0}_{e_p} \ket{0}_{e_q} \ket{0}_{\zeta} { \ket{1} }_{me}  \ket{y_P(\mathcal{F},j)}_{b_p} \ket{y_Q(\mathcal{F},j)}_{b_q} \nonumber \\
		+ & \frac{1}{\sqrt{\mathcal{D}}} \sum _{j=0}^{\mathcal{D}-1}  e^{i\theta _{j}}\ { \zeta _{\mathcal{F}_j}^{\mathcal{F}}  } {\rho _{j} } \ket{j }_{id}  \ket{\mathcal{F}_{j}}_s \ket{\mathcal{F}}_{cp} \ket{0}_{e_p} \ket{0} _{e_q}  \ket{0} _{\zeta} {  \ket{0} }_{me}  \ket{y_P(\mathcal{F},j)}_{b_p} \ket{y_Q(\mathcal{F},j)}_{b_q} . 
		\label{eq:mid23}
	\end{align}
	
	We substitute Eq. \eqref{eq:mid23} to Eq. \eqref{eq:mid20} to have
	\begin{align}
		&	( \langle \mathcal{G} |_s \otimes \langle 0| _a ) \big( \mathcal{T}^{\dagger}_b {S} \mathcal{T}_f \big) ( | \mathcal{F} \rangle _s \otimes \ket{0} _a )  \nonumber \\ 
		= & \frac{1}{\mathcal{D}}  \Bigg\{  \sum _{k=0}^{\mathcal{D}-1} \ket{k}_{id} \ket{\mathcal{G}}_{s} \ket{\mathcal{G}_{k}}_{cp} \ket{y_P(\mathcal{G},k)}_{e_p} \ket{y_Q(\mathcal{G},k)}_{e_q}  \ket{0} _{\zeta} \ket{0} _{me}    \ket{0}_{b_p}  \ket{0}_{b_q}  \Bigg\} ^{\dagger} \nonumber \\
		& \ \ \ \Bigg\{  \sum _{j=0}^{\mathcal{D}-1}  e^{i\theta _{j}} \zeta _{\mathcal{F}_j}^{\mathcal{F}} \ {\rho _{j} } \ket{j}_{id}  \ket{\mathcal{F}_{j}}_s \ket{\mathcal{F}}_{cp} \ket{0}_{e_p} \ket{0}_{e_q}  \ket{0}_{\zeta} \ket{0}_{me}  \ket{y_P(\mathcal{F},j)}_{b_p} \ket{y_Q(\mathcal{F},j)}_{b_q} 
		\Bigg\} \nonumber \\
		= & \frac{1}{\mathcal{D}}  \sum _{k=0}^{\mathcal{D}-1} \sum _{j=0}^{\mathcal{D}-1}  \zeta _{\mathcal{F}_j}^{\mathcal{F}} \  {\rho _{j} } e^{i\theta _{j}} \delta _{ j,k} \ \delta _{\mathcal{G}, \mathcal{F}_j } \ \delta _{\mathcal{G}_k , \mathcal{F}} \ \delta _{y_P(\mathcal{G},k) ,0 } \ \delta _{y_Q(\mathcal{G},k) ,0 } \ \delta _{y_P(\mathcal{F},j) ,0 } \ \delta _{y_Q(\mathcal{F},j) ,0 }     \nonumber \\
		=& \frac{1}{\mathcal{D}} \sum _{j=0} ^{\mathcal{D}-1}  \zeta _{\mathcal{F}_j}^{\mathcal{F}} \  \rho _{j} e^{i\theta _{j}} \delta _{\mathcal{G}, \mathcal{F}_j } \ \delta _{\mathcal{G}_j , \mathcal{F}} \ \delta _{y_P(\mathcal{G}, j) ,0 } \ \delta _{y_Q(\mathcal{G},j) ,0 } \ \delta _{y_P(\mathcal{F},j) ,0 } \ \delta _{y_Q(\mathcal{F},j) ,0 }  \nonumber \\
		=& \frac{1}{\mathcal{D} \Lambda _m} \sum _{j=0} ^{\mathcal{D}-1}  \langle \mathcal{G} | b^{\dagger}_{Q_j} b_{P_j} | \mathcal{F} \rangle  \langle Q_j |H | P_j \rangle \delta _{\mathcal{G}, \mathcal{F}_j } \ \delta _{\mathcal{G}_j , \mathcal{F}} \ \delta _{y_P(\mathcal{G}, j) ,0 } \ \delta _{y_Q(\mathcal{G},j) ,0 } \ \delta _{y_P(\mathcal{F},j) ,0 } \ \delta _{y_Q(\mathcal{F},j) ,0 } .
	\end{align}
	The right hand side of the above equation presents all the contribution of the scaled few-body matrix elements (in the form of  $ \rho _{j} e^{i\theta _{j}} $) weighted by corresponding factors ($ \zeta _{\mathcal{F}_j}^{\mathcal{F}} = \langle \mathcal{G} | b^{\dagger}_{Q_j} b_{P_j} | \mathcal{F} \rangle = \pm 1$) due to the anticommutation relations. 
	The delta functions guarantee that only the matrix elements, which are allowed by the Pauli exclusion principle, are counted into the many-body matrix element $ \langle \mathcal{G} | H' | \mathcal{F} \rangle $. 
	
	This completes the proof of Eq. \eqref{eq:BE_Hamiltonian}.
	
\end{proof}

\section{The framework of SA-QKSD}
\label{sec:SA_QKSD_framework}

We provide the Hamiltonian input scheme in Eq. \eqref{eq:BE_Hamiltonian}. Based on this input scheme, we can block encode the Chebyshev polynomial of the first kind \cite{arfken2013mathematical} as a function of $H' = H/(\mathcal{D}\Lambda)$ with the qubitization \cite{Low2019hamiltonian,gilyen2019quantum,PRXQuantum.2.040203} as 
\begin{align}
	\langle \psi_0 | T_{2k+1}(H') | \psi_0 \rangle  
	=& 	( \langle {\psi _0} |_s \otimes \langle 0| _a ) \big[  U_H {\Pi} ( U^{\dagger}_H {\Pi} U_H {\Pi})^k \big] ( | {\psi _0}  \rangle _s \otimes \ket{0} _a ) \label{eq:Chebyshev_odd} , \\
	\langle \psi_0 | T_{2k}(H')  | \psi_0 \rangle  
	=&( \langle \psi_0 |_s \otimes \langle 0| _a )  ( U^{\dagger}_H {\Pi} U_H {\Pi})^k  ( | \psi_0 \rangle _s \otimes \ket{0} _a ) ,
	\label{eq:Chebyshev_even}
\end{align}
with $k = 0, 1, 2, \cdots $. The pivot Fock state is $\ket{ \psi _0 }$. The reflection operator  $\Pi \equiv (2 \ket{0}_a \bra{0}_a - \mathds{I}_a )\otimes \mathds{I}_s$ produces the reflection around $\ket{0}_a$. The circuit to input  $ T_j(H') $ can be constructed by alternative applications of $U_H\Pi $ or $ U_H^{\dagger } \Pi $ \cite{lin2022lecture}. 

With the input scheme of the Chebyshev polynomial $T_i(H')$, we can construct the Krylov bases based on the pivot Fock state $\ket{ \psi _0 }$ as \cite{Kirby_2023}
\begin{equation}
	\mathcal{B}_{\mathcal{K}} = \{ \ket{\psi _i} = T_k (H') \ket{\psi _0} | k = 0, 1, 2,  \cdots , \mathcal{K} -1  \} .
	\label{eq:B_krylov}
\end{equation}
Compared to the typical Krylov basis set that are constructed from the actions of the Hamiltonian, i.e.,
\begin{equation}
	\{ \ket{\phi _k}  = H^k \ket{\psi _0} | k = 0, 1, 2,  \cdots , \mathcal{K} - 1 \}, 
\end{equation}
the basis set $\mathcal{B}_{\mathcal{K}} $ span the same Hilbert space spanned by the set  $ \{ \ket{\phi _k} \} $. 

We note that our input scheme [Eq. \eqref{eq:BE_Hamiltonian}] respects the symmetries of the Hamiltonian $H$. 
The input scheme of the Chebyshev polynomials [Eqs. (\ref{eq:Chebyshev_odd}, \ref{eq:Chebyshev_even})] also respects the symmetries of $H$.
Working in the second-quantized formalism, it is straightforward to prepare the pivot Fock state $\ket{\psi _0}$ to have specific symmetries of the many-fermion system. As such, we can prepare the Krylov bases $\mathcal{B}_{\mathcal{K}} $ with elected symmetries, which span the SA Hilbert subspace in which we can solve for the eigenenergy of the lowest state with the elected symmetry. This idea is useful for accessing the excited spectrum that can not be reached by standard QKSD methods \cite{Huggins_2020,doi:10.1021/acs.jctc.9b01125,PhysRevA.105.022417,PRXQuantum.3.020323,doi:10.1137/21M145954X,PhysRevA.107.032414,Kirby_2023} for structure calculations: with the symmetry regulations, it is possible to exclude those low-lying states from the SA Krylov subspace by manipulating the symmetries.\footnote{As an example of the symmetry regulation shown in the main text, the Hamiltonian of the many-nucleon system preserves the particle number and the projection $M_J$ of the total angular momentum $J$ of the system, it is feasible to generate the set $\mathcal{B}_{\mathcal{K}}$ with specific $M_J$ by electing $\ket{\psi _0}$ to be of fixed $M_J$. With this $M_J$ regulation, those eigenstates with $J < M_J$ are naturally excluded in the spectrum of the many-nucleon Hamiltonian in the SA Hilbert subspace spanned by $\mathcal{B}_{\mathcal{K}}$.}  

With the SA Krylov basis set $\mathcal{B}_{\mathcal{K}} $, we can construct the Hamiltonian matrix element as
\begin{equation}
	H'_{ij} = \langle \psi _i | H' | \psi _j \rangle = \langle \psi _0 | T_i (H') H' T_j (H') | \psi _0 \rangle ,
\end{equation}
and the overlap matrix element as
\begin{equation}
	\Upsilon _{ij} = \langle \psi _i | \psi _j \rangle = \langle \psi _0 | T_i (H') T_j (H') | \psi _0 \rangle .
\end{equation}
These matrix elements can be simplified utilizing the following relations \cite{arfken2013mathematical}
\begin{equation}
	T_i(x)T_j(x) = \frac{1}{2} \big( T_{i+j}(x) + T_{|i-j|} (x) \big) ,\ T_1(H') = H', \ T_0(H') = 1.
\end{equation}
After simplifications, the elements $H'_{ij}$ and $\Upsilon _{ij}$ can be expressed in terms of the expectations $ \langle T_k(H') \rangle _0 \equiv  \bra{\psi _0} T_k(H') \ket{\psi_0} $ as
\begin{align}
	H'_{ij} =& \frac{1}{4} \big( \langle T_{i+j+1}(H') \rangle _0 + \langle T_{|i+j-1|}(H') \rangle _0 + \langle T_{|i-j+1|}(H') \rangle _0  + \langle T_{|i-j-1|}(H') \rangle _0 \big) \label{eq:H_ij} \\
	\Upsilon _{ij} =&  \frac{1}{ 2 } \big( \langle T_{i+j}(H') \rangle _0 + \langle T_{|i-j|}(H') \rangle _0 \big) .
	\label{eq:S_ij}
\end{align}

In general, the expectations $ \{ \langle T_k(H') \rangle _0  \} $ of quantum many-body systems are computationally challenging on classical computers.
Quantum computing techniques can offer a promising path to address such classically hard problem. For example, one can evaluate these expectations based on the input model for the function $T_k(H')$ introduced above with the implementation of the Hadamard test \cite{nielsen2010quantum} or other methods such as the multifidelity estimation protocol \cite{PhysRevA.105.022417}. 

The dimension of the SA Krylov subspace for evaluating the eigenenergy of the lowest state is expected to be greatly reduced from the original many-body Hamiltonian in the SP basis. 
A general bound on $\mathcal{K}$ [Eq. \eqref{eq:B_krylov}] to compute the ground-state energy within the error $\mathcal{E}$ is \cite{Kirby_2023} 
\begin{equation}
	\mathcal{K} = \Theta [  ( \log | \gamma _0 |^{-1}  + \log {\mathcal{E}}^{-1}  ) \cdot \min ( {\mathcal{E}}^{-1}, {\Delta}^{-1} )  ] ,
	\label{eq:dim_or_KrylovBasisSpace}
\end{equation}
with $\Delta $ being the spectral gap. $ \gamma_0 $ denotes the overlap between the pivot $\ket{\psi_0 }$ and the ground state of $H'$ in the SA Hilbert subspace. 

With the expectation values $  \{\langle T_k(H') \rangle _0 \} $ evaluated on quantum computers, we access the matrix elements of $ H'_{ij} $ and $ \Upsilon _{ij}$ in the SA Krylov basis $ \{\ket{\psi _i} \} $ according to Eqs. \eqref{eq:H_ij} and \eqref{eq:S_ij}. We can then construct the generalized eigenvalue problem 
\begin{equation}
	H' \ket{\Psi } = \lambda \Upsilon \ket{\Psi } .
	\label{eq:generalizedEvalProblem}
\end{equation}
It is noteworthy that the overlap matrix can be of large condition number due to the linear dependency of the Krylov basis. This problem can be potentially relieved by the multireference approach \cite{doi:10.1021/acs.jctc.9b01125} where the Krylov basis is constructed from an elected initial set of orthogonal reference pivot states according to a measurement based identification of the most important determinants in a simple trial wave function. 

As a straightforward approach in solving the generalized eigenvalue problem [Eq. \eqref{eq:generalizedEvalProblem}], we apply the canonical orthogonalization (CO) method \cite{LOWDIN1970,jiao2015application,PRXQuantum.2.010333}. In particular, the overlap matrix $\Upsilon$ admits the diagonalization with a unitary matrix $V$ such that
\begin{equation}
	V^{\dagger} \Upsilon V =  \upsilon ,
\end{equation}
where $ \upsilon $ is the diagonal matrix with the diagonal entries being the eigenvalues of $\Upsilon $. These diagonal entries are all positive since $\Upsilon $ is positive definite. With the transformation matrix defined as $ W = V \upsilon ^{-1/2} $, the generalized eigenvalue problem [Eq. \eqref{eq:generalizedEvalProblem}] can be transformed into a standard eigenvalue problem 
\begin{equation}
	H'' \ket{\Psi '} = \lambda \ket{\Psi '} 
	\label{eq:evalue_problem}
\end{equation}
with $H'' = W^{\dagger} H' W$ and $\ket{\Psi '} = W^{-1} \ket{\Psi } $. By solving the eigenvalue problem, one obtains the eigenvalues and eigenvectors. 

In practical applications, the matrix $H'$ and $ \Upsilon $ is nearly singular when the Krylov basis is close to linearly dependent. Based on the diagonalization of overlap matrix $\Upsilon$, one can apply the minimum eigenvalues of $\Upsilon$ to measure the linear dependency of the eigenvectors $\{ \ket{\Psi} \}$. 
According to the CO method, one introduces a threshold parameter $\xi$, where one retains only the eigenvectors of $\Upsilon$ of which the corresponding eigenvalues are larger than $\xi$. These elected eigenvectors and eigenvalues of $\Upsilon $ are employed to construct a unitary matrix $\widetilde{V}$, which is of smaller size than $V$, and the diagonal matrix $\widetilde{\upsilon}$. The transformation matrix in practical calculations is then defined as $\widetilde{W} = \widetilde{V} \widetilde{\upsilon} ^{1/2}$. As such, one empirically finds the set of eigenstates $\{ \ket{\Psi '} \}$ that are maximally linearly independent. 

Though there lacks a general theory of why the CO method works \cite{doi:10.1137/21M145954X}, the CO method is capable of approximating the lowest eigenvalue of $H'$ as a function of the truncation parameter $\xi$, especially when the overlap matrix becomes ill conditioned. In some of our model problems where the generalized eigenvalue problems are of limited size, we find that the CO method also manages to approximate the eigenenergies of excited states in addition to the ground-state energy.

\section{Two-nucleon matrix elements}
\label{sec:numerical_calculations_info}

\subsection{The oscillator basis}

We adopt the harmonic oscillator basis in this work \cite{suhonen2007nucleons}. In the coordinate representation, we can write the harmonic oscillator basis as \cite{Du:2018tce,Du:2022zds}
\begin{align}
	\langle \vec{r} | nlsjm_j t \tau  \rangle = R_{nl}(r) \sum _{m_l } \sum _{m_s} (l m_l sm_s|j m_j) Y_{lm_l}(\Omega _{\hat{r}}  ) \chi _{sm_s} \chi _{t \tau},
	\label{eq:HO_basis_function}
\end{align}
with the radial quantum number $n$, the orbital angular momentum $l$, the spin $s$. The total angular momentum $j$ is coupled from $l$ and $s$. The projections of $l$, $s$, and $j$ are denoted as $m_l$, $m_s$, and $m_j$, respectively. 
$\chi _{sm_s} $ denotes the spinor with the spin $s = 1/2$ for fermions and spin projection $m_s=\pm 1/2$. $\chi _{t \tau} $ denotes the isospinor with $t=1/2$ for the nucleons (proton and neutron) and its projection $\tau$ being $+1/2$ for the proton and $-1/2$ for the neutron. $ Y_{lm_l}( \Omega _{\hat{r}} ) $ denotes the spherical harmonics \cite{arfken2013mathematical} with $\Omega _{\hat{r}}$ being the solid angle. The Clebsch-Gordan coefficient $ (l m_l sm_s|j m_j) $ follows the Condon-Shortley convention \cite{suhonen2007nucleons}.

The radial part of the wave function is \cite{Du:2018tce,Du:2022zds}
\begin{align}
	R_{nl}(r) = \sqrt{ \frac{2n!}{r_0^3 \Gamma (n+l+\frac{3}{2}) } } \Big( \frac{r}{r_0} \Big)^l \exp \Big[ - \frac{r^2}{2r_0^2} \Big] L_n^{l+\frac{1}{2}} \Big( \frac{r^2}{r_0^2} \Big) ,
\end{align}
where $\Gamma (\cdot)$ is the Gamma function and $ L_n^{\alpha } (\cdot) $ is the associated Laguerre polynomial \cite{arfken2013mathematical}. $ r_0 = 1/\sqrt{m_N \omega} $ denotes the oscillator length, with $ m_N $ being the nucleon mass (taken to be 938.919 MeV in this work) and $\omega $ being the oscillator strength (taken to be 12 MeV in this work to resemble the nuclear environment of the Calcium isotopes).

\subsection{The quadrupole operator in the harmonic oscillator basis}

The matrix element of the quadrupole operator in the harmonic oscillator basis is
\begin{align}
	& \langle n_{p} l_{p} s_{p} j_{p} m_{p} t _p \tau _p | r^2 Y_{2 \sigma } | n_{q} l_{q} s_{q} j_{q} m_{q} t_q \tau _q \rangle \nonumber \\
	=& \int R^{\ast} _{n_{p} l_{p} }(r) r^2 R_{n_{q l_{q}} }(r) \ r^2 dr \times \Bigg[ \sum _{m_{l_{p}} ,  m_{s_{p}}} \sum _{m_{l_{q}} ,  m_{s_{q }}}  (l_{p } m_{l_{p}} s_{p } m_{s_{p}} | j_{p } m_{p} )^{\ast} (l_{q } m_{l_{q}} s_{q } m_{s_{q}} | j_{q } m_{q} ) \nonumber \\
	& \times \int Y^{\ast}_{l_{p} m_{l_{p}} } (\Omega ) Y_{2\sigma} (\Omega _{\hat{r}}) Y_{l_{q} m_{l_{q}} }(\Omega _{\hat{r}}) d \Omega _{\hat{r}} \ 
	\chi _{s_{p}m_{s_{p}}}^{\dag} \chi _{s_{q}m_{s_{q}}} \chi ^{\dagger} _{t_p \tau _p} \chi _{t_q \tau _q} \Bigg] .
\end{align}

The radial part of the integral yields
\begin{align}
	&  \int R^{\ast} _{n_{p} l_{p} }(r) r^2 R_{n_{q l_{q}} }(r) \ r^2 dr \nonumber \\
	=\ \ & r_0^2 \Bigg[(2n_{p }+l_{p} +\frac{3}{2})\delta_{n_{p}n_{q}}-\sqrt{(n_{p}+l_{p}+\frac{3}{2})(n_{p}+1)}\delta_{n_{p},n_{q}-1}-\sqrt{(n_{q}+l_{q}+\frac{3}{2})(n_{q }+1)}\delta_{n_{p},n_{q }+1} \Bigg] \delta_{l_{p},l_{q }} \nonumber \\
	+ & r_0^2 \Bigg[\sqrt{(n_{q }+l_{p } +\frac{3}{2})(n_{q }+l_{p}+\frac{5}{2})}\delta_{n_{p} n_{q}}-2\sqrt{(n_{q}+1)(n_{q}+l_{p} +\frac{5}{2})}\delta_{n_{p},n_{q}+1}+\sqrt{(n_{q}+1)(n_{q}+2)}\delta_{n_{p},n_{q}+2} \Bigg] \delta _{l_{q},l_{p}+2} \nonumber \\
	+ & r_0^2 \Bigg[\sqrt{(n_{p}+l_{q}+\frac{3}{2})(n_{p}+l_{q}+\frac{5}{2})}\delta_{n_{p} n_{q}}-2\sqrt{(n_{p}+1)(n_{p}+l_{q}+\frac{5}{2})}\delta_{n_{q},n_{p}+1}+\sqrt{(n_{p}+1)(n_{p}+2)}\delta_{n_{q},n_{p}+2} \Bigg] \delta _{l_{p},l_{q}+2}  .
\end{align}

The integral of three harmonic oscillator functions can be evaluated via the {\it Gaunt formula} \cite{arfken2013mathematical} as
\begin{align}
	\int Y^{\ast}_{l_{p} m_{l_{p}} } (\Omega _{\hat{r}}) Y_{2\sigma} (\Omega _{\hat{r}}) Y_{l_{q} m_{l_{q}} }(\Omega _{\hat{r}}) d \Omega _{\hat{r}}  = 
	(-1)^{m_{l_{p}}} \frac{\hat{l}_{p} \hat{ 2 }  \hat{ l }_{q} }{ \sqrt{4\pi} } 
	\begin{pmatrix}
		l_{p} & 2 & l_{q}\\ 
		0 & 0 & 0
	\end{pmatrix}
	\begin{pmatrix}
		l_{p} & 2 & l_{q}\\ 
		-m_{l_{p}} & \sigma & m_{l_{q}}
	\end{pmatrix}  ,
\end{align}
with $\hat{l}_{p} \equiv \sqrt{2l_{p} +1 }$ and 
\begin{equation}
	\begin{pmatrix}
		l_{p} & 2 & l_{q}\\ 
		0 & 0 & 0
	\end{pmatrix}, \ 
	\begin{pmatrix}
		l_{p} & 2 & l_{q}\\ 
		-m_{l_{p}} & \sigma & m_{l_{q}}
	\end{pmatrix} 
	\nonumber
\end{equation}
are the the 3$j$ symbol \cite{suhonen2007nucleons}.

The products of the spinors and isospinors yield the delta functions
\begin{align}
	\chi _{s_{p}m_{s_{p}}}^{\dag} \chi _{s_{q}m_{s_{q}}}  = \delta _{s_{p}, s_{q}} \delta _{m_{s_{p}}, m_{s_{q }} }  , \ \chi ^{\dagger} _{t_p \tau _p} \chi _{t_q \tau _q} = \delta _{t_p , t_q} \delta _{\tau _p , \tau _q} .
\end{align}

Based on the oscillator basis [Eq. \eqref{eq:HO_basis_function}], we can evaluate the two-nucleon matrix element of the pairing plus quadrupole-quadrupole Hamiltonian [Eq. (11) in the main text] as
\begin{align}
	\langle pq | H_A  | u v \rangle = \frac{1}{2} \int  \Big[
	\phi _p^{\dag}(1) \phi _q^{\dag}(2) -(p \leftrightarrow q)
	\Big] H_A \Big[
	\phi _u(1) \phi _v(2) -(u \leftrightarrow v)
	\Big] d\vec{x}_1 d\vec{x}_2 ,
\end{align}
where $\phi _{i}$ denotes the harmonic oscillator wave function of the ${i}$th SP basis state that is specified by the spatial, spin, and isospin quantum numbers. The labels ``$1$" and ``$2$" in the parentheses are the nucleon indices. The integral $ \int d\vec{x}_1 d\vec{x}_2 $ denotes the integration over the spatial variables and the summation over the spin and isospin variables. 

In the $0f_{7/2}$ valance space (Table \ref{tab:SPbasis_0f}), the two-body matrix elements of $H_A$ are computed on the classical computer; these elements are presented in Table \ref{tab:ME_table}. In practice, a limited number of such two-body matrix elements are present to construct the complicated many-body matrix element in the full configuration interaction calculations.

\begin{table*}[!ht] 
	\centering
	\caption{The SP bases in the $ 0f_{7/2} $ valance space. The quantum numbers of each SP basis are presented, whereas the spin and isospin quantum numbers are understood to be ${1}/{2}$. Each SP basis is indexed and mapped to a distinct qubit (second column).}
	\label{tab:SPbasis_0f}
	\begin{tabular}{ccccccc}
		\hline \hline
		\multicolumn{1}{l}{}        & SP basis (qubit) & $\ n\ $ & $\ l \ $ & $\ 2j \ $ & $\ 2m_j\ $ & $\ 2\tau \ $ \\ \hline 
		\multirow{6}{*}{$0f_{7/2}$} & {\bf 0}                & $0$   & $3$   & $7$    & $+7$     & $-1$     \\
		& {\bf 1}                & $0 $  & $3$   & $7$    & $-7$     & $-1$      \\
		& {\bf 2}                & $0$   & $3$   & $7$    & $+5$     & $-1$      \\
		& {\bf 3}                & $0 $  & $3$   & $7$    & $-5$     & $-1$      \\
		& {\bf 4}                & $0 $  & $3$   & $7$    & $+3$     & $-1$      \\
		& {\bf 5}                & $0 $  & $3$   & $7$    & $-3$     & $-1$      \\ 
		& {\bf 6}                & $0 $  & $3$   & $7$    & $+1$     & $-1$      \\
		& {\bf 7}                & $0 $  & $3$   & $7$    & $-1$     & $-1$      \\
		\hline \hline
	\end{tabular}
\end{table*}

\begin{table*}[ht!] 
	\centering
	\caption{Two-body matrix element of $H_A$ [Eq. (11) in the main text] in the valance space $0f_{7/2}$. The index $i$ corresponds to a specific set $\{ p_i, q_i, u_i, v_i  \}$, where each of the indices $ p_i, q_i, u_i$, and $v_i $ corresponds to a SP basis (and also a distinct qubit) according to the scheme shown in Table \ref{tab:SPbasis_0f}. All the matrix elements are in units of MeVs.}
	\begin{tabular}{c|ccccc | c|ccccc | c|ccccc | c|ccccc }
		\hline \hline
		$i$ & $p_i$ & $q_i$ & $u_i$ & $v_i$ & $\langle p_i q_i |H_A |u_i v_i  \rangle $ 
		& $i$ & $p_i$ & $q_i$ & $u_i$ & $v_i$ & $\langle p_i q_i |H_A |u_i v_i  \rangle $ 
		& $i$ & $p_i$ & $q_i$ & $u_i$ & $v_i$ & $\langle p_i q_i |H_A |u_i v_i  \rangle $
		& $i$ & $p_i$ & $q_i$ & $u_i$ & $v_i$ & $\langle p_i q_i |H_A |u_i v_i  \rangle $ \\
		\hline
		0 & 0 & 1 & 0 & 1 & -0.965525  &  16 & 1 & 2 & 5 & 6 & -0.342159 & 32 & 2 & 6 & 2 & 6 & 0.584347 &  48 & 4 & 6 & 0 & 5 & -0.395091    \\
		1 & 0 & 1 & 2 & 3 & 0.848655   &  17 & 1 & 3 & 1 & 3 & 0.584347  & 33 & 2 & 7 & 0 & 5 & 0.342159 &  49 & 4 & 6 & 4 & 6 & -0.0834782   \\
		2 & 0 & 1 & 4 & 5 & -0.381178  &  18 & 1 & 4 & 1 & 4 & 0.350608   & 34 & 2 & 7 & 4 & 6 & 0.578354  &  50 & 4 & 6 & 2 & 7 & 0.578354     \\
		3 & 0 & 1 & 6 & 7 & 0.147439   &  19 & 1 & 4 & 3 & 6 & 0.342159  & 35 & 2 & 7 & 2 & 7 & 0.0834782 &  51 & 4 & 7 & 0 & 3 & -0.342159   \\
		4 & 0 & 2 & 0 & 2 & 0.584347   &  20 & 1 & 4 & 5 & 7 & -0.395091  & 36 & 3 & 4 & 1 & 2 & 0.612072 &  52 & 4 & 7 & 2 & 5 & 0.298661    \\
		5 & 0 & 3 & 0 & 3 & -0.116869  &  21 & 1 & 5 & 1 & 5 & 0.584347  & 37 & 3 & 4 & 3 & 4 & 0.0500869 &  53 & 4 & 7 & 4 & 7 & 0.417391    \\
		6 & 0 & 3 & 2 & 5 & 0.612072   &  22 & 1 & 6 & 1 & 6 & 0.584347  & 38 & 3 & 4 & 5 & 6 & 0.298661 &  54 & 5 & 6 & 1 & 2 & -0.342159   \\
		7 & 0 & 3 & 4 & 7 & -0.342159  &  23 & 1 & 7 & 1 & 7 & 0.584347  & 39 & 3 & 5 & 3 & 5 & 0.584347 &  55 & 5 & 6 & 3 & 4 & 0.298661    \\
		8 & 0 & 4 & 0 & 4 & 0.584347   &  24 & 2 & 3 & 0 & 1 & 0.848655  & 40 & 3 & 6 & 1 & 4 & 0.342159 &  56 & 5 & 6 & 5 & 6 & 0.417391    \\
		9 & 0 & 5 & 0 & 5 & 0.350608   &  25 & 2 & 3 & 2 & 3 & -0.164134 & 41 & 3 & 6 & 3 & 6 & 0.0834782 &  57 & 5 & 7 & 1 & 4 & -0.395091    \\
		10 & 0 & 5 & 4 & 6 & -0.395091 &  26 & 2 & 3 & 4 & 5 & 0.681699  & 42 & 3 & 6 & 5 & 7 & 0.578354  &  58 & 5 & 7 & 3 & 6 & 0.578354     \\
		11 & 0 & 5 & 2 & 7 & 0.342159  &  27 & 2 & 3 & 6 & 7 & -0.648308 & 43 & 3 & 7 & 3 & 7 & 0.584347 &  59 & 5 & 7 & 5 & 7 & -0.0834782   \\
		12 & 0 & 6 & 0 & 6 & 0.584347  &  28 & 2 & 4 & 2 & 4 & 0.584347  & 44 & 4 & 5 & 0 & 1 & -0.381178 &  60 & 6 & 7 & 0 & 1 & 0.147439    \\
		13 & 0 & 7 & 0 & 7 & 0.584347  &  29 & 2 & 5 & 0 & 3 & 0.612072  & 45 & 4 & 5 & 2 & 3 & 0.681699 &  61 & 6 & 7 & 2 & 3 & -0.648308   \\
		14 & 1 & 2 & 1 & 2 & -0.116869  & 30 & 2 & 5 & 2 & 5 & 0.0500869  & 46 & 4 & 5 & 4 & 5 & -0.297699 & 62 & 6 & 7 & 4 & 5 & 0.982221    \\
		15 & 1 & 2 & 3 & 4 & 0.612072   & 31 & 2 & 5 & 4 & 7 & 0.298660  & 47 & 4 & 5 & 6 & 7 & 0.982221  & 63 & 6 & 7 & 6 & 7 & -0.564830   \\
		\hline \hline
	\end{tabular}
	\label{tab:ME_table}
\end{table*}

\bibliography{apssamp}

\end{document}